\documentclass{aa}
\usepackage{graphicx}

\begin{document}

\newcommand{\Msun}{\mbox{$M_{\odot}\;$}}
\def\la{\;\raise0.3ex\hbox{$<$\kern-0.75em\raise-1.1ex\hbox{$\sim$}}\;}
\def\ga{\;\raise0.3ex\hbox{$>$\kern-0.75em\raise-1.1ex\hbox{$\sim$}}\;}

\authorrunning{D.~Konenkov, U.~Geppert}
\titlerunning{Residual Magnetic Fields in Millisecond Pulsars}

\title{On the Nature of the Residual Magnetic Fields in Millisecond Pulsars}

\author{D.~Konenkov\inst{1,2} \and U.~Geppert\inst{2}}
\institute{A.F.Ioffe Institute of Physics and Technology,
           Politechnicheskaya 26, 194021 St.Petersburg, Russia;
           dyk@astro.ioffe.rssi.ru \\
           \and
           Astrophysikalisches Institut Potsdam,
           An der Sternwarte 16, 14482 Potsdam, Germany;
           urme@aip.de \\
           }

 \date{Received date ; accepted date}

\abstract{We consider the expulsion of proton fluxoids along
neutron vortices from the superfluid/superconductive core of
neutron star with weak ($B<10^{10}$ G) magnetic field. The
velocity of fluxoids is calculated from the balance of buoyancy,
drag and crustal forces. We show, that the proton fluxoids can
leave the superfluid core sliding {\it along} the neutron
vortices on a timescale of about $10^7$ years. An alternative
possibility is that fluxoids are aligned with the vortices on the
same timescale. As the result, non--aligned surface  magnetic
fields of millisecond pulsars can be sustained for $\ga 10^9$
years only in case of a comparable dissipation timescale of the
currents in the neutron star crust. This defines upper limits of
the impurity concentration in the neutron star crust: $Q \la 0.1$
if a stiff equation of state determines the density profile.
}

\maketitle

\section{Introduction}

There are observational evidences that the magnetic fields (MFs)
of millisecond pulsars (MSPs) are stationary over the age of $\ga
10^9$ years. The spin-down ages of many  MSPs and the cooling
ages of their white dwarf companions exceed $10^9$ years
(Kulkarni 1986; Danziger et al. 1993; Kulkarni et al. 1991; Bell
et al. 1995). MSP statistics indicates also, that the ages of MSPs
must be greater than $10^9$ years (Bhattacharya \& Srinivasan
1991 and references therein).

\noindent We assume that both protons and neutrons are in a
superfluid state in the whole core (Baym et al. 1969). The softer
the EOS and the more massive the NS, the larger is the probability
that protons and neutrons in the central region of the core  are
in a normal state (Page 1998a), which implies a completely other
core field evolution (see e.g. Goldreich \& Reisenegger 1992). On
the other hand, the millisecond pulsars are very old NSs. Their
internal temperature is expected to be much less than any
non--zero superfluid transition temperature (Page 2001, private
communication), favouring the superfluid state in the entire core.

\noindent The  superfluid phase transition takes place early in
NS's life (Page 1998b) and the MF penetrates the core as an array
of proton flux tubes (fluxoids), each of them  carrying a quantum
of the magnetic flux $\Phi_0=hc/2e \approx 2\cdot 10^{-7}$  G
cm$^2$. Inside the fluxoids the MF is as high as $B_{\rm p} \sim
10^{15}$ G, and it decays outwardly exponentially with the London
penetration depth $\lambda \sim 10^{-11}$ cm. The core superfluid
rotates forming a discrete array of neutron vortices. The kernels
of vortices and fluxoids  consist of normal matter, their radii
correspond to the  coherence lengths $\xi_{\rm n}$ and $\xi_{\rm
p}$, respectively.

\noindent Because of a superfluid drag effect the neutron
vortices are also magnetized (Alpar et al. 1984). The MF inside
each vortex is $B_{\rm n} \sim 10^{15}$ G. When a vortex crosses a
fluxoid, the interaction of their MF results in a potential
barrier, $E_{\rm m} \sim 10$ MeV per intersection (Ding et al.
1993, DCC hereafter), corresponding to a pinning force $F_{\rm m}
\sim E_{\rm m}/\lambda \sim 10^6$ dyn per intersection.

\noindent The interaction energy due to density perturbations in
the  center of a fluxoid is $E_{\rm p} \sim 0.1-1$ MeV per
intersection and causes a pinning force $F_{\rm p} \sim E_{\rm p}
/ \xi_{\rm n}$, which is in the same order of magnitude as
$F_{\rm m}$ (Sauls 1989).

\noindent The strong interaction between neutron vortices and
proton fluxoids led many authors to develop models which consider
the MF evolution together with the NS spin--down. Srinivasan et
al. (1990) and Jahan--Miri \& Bhattacharya (1994) calculated the
expulsion of the magnetic flux from the core  assuming that
fluxoids and vortices always move with the same velocity. DCC and
Jahan--Miri (2000) calculated the magnetic evolution in more
realistic models, in which other forces, acting on the fluxoids
(buoyancy, drag, tension) were taken into consideration. Konenkov
\& Geppert (2000, 2001) (KG00, KG01 hereafter) extended these
models by taking into account the back--reaction of the crust
(``crustal'' forces) onto the rate of flux expulsion. All  these
authors reported the occurrence of a long--living ($\sim 10^{10}$
years) low  MF component, which could be responsible for a
non--decaying ``residual''  MF of MSPs.

\noindent In the present letter we calculate the flux expulsion
timescale in the core of MSPs when the fluxoids slide {\it along}
(parallel) to the neutron vortices, i.e., when the interpinning
between vortices and fluxoids  can be neglected. This possibility
was first mentioned by Muslimov \& Tsygan (1985). We show, that
this timescale is about $10^7$ years.

\section{Description of the model}

Besides the force  exerted by the neutron vortices, there exist
other forces which act onto the fluxoids in the NS core. The
buoyancy force, acting per unit length of the fluxoid, is given
by (Muslimov \& Tsygan, 1985):
\begin{equation}
\vec{f_{\rm b}} = \left(\frac{\Phi_0}{4\pi \lambda}\right)^2
      \frac{1}{R_{\rm c}}\ln\left (\frac{\lambda}{\xi_{\rm p}}\right) ~\vec{e}_r ,
\label{f_b_eq}
\end{equation}
where $R_{\rm c}$ is the  radius of the NS core and $\vec{e}_r$ is
the unit vector in radial direction when spherical coordinates are
used. The buoyancy force acts always radially outward.

\noindent The drag  force, which arises due to the scattering of
the relativistic electrons on the MF of the fluxoid, is given by
(Harvey et al. 1986; Jones 1987):
\begin{equation}
\vec{f_{\rm v}} = -\frac{3\pi}{64}\frac{n_{\rm e} e^2
\Phi_0^2}{E_{\rm F} \lambda}\frac{\vec{v}_{\rm p}} {c} ,
\label{f_v_eq}
\end{equation}
where $n_{\rm e}$ and $E_{\rm F}$ are the number density and the
Fermi energy of electrons, $c$  is the speed of light, and
$\vec{v}_{\rm p}$ the velocity of the fluxoid's motion through the
core. The drag force acts opposite to the velocity of the
fluxoid. This equation is valid only if collective effects are
ignored (see DCC).

\noindent In MSPs the proton fluxoids are indeed expected to be
unable to cut through the neutron vortices, and the motion of
arbitrarily oriented fluxoids in the direction perpendicular to
the rotational axis occurs only on the spin--down timescale of
MSPs. However, the displacement of the fluxoid parallel to the
neutron vortex  does not change the pinning energy, thus the
movement of the fluxoids parallel  to the vortices is not
restricted by the pinning force. There  is a component of the
buoyancy force, $\vec{f}_{\rm b||}$, which is not compensated
locally  by the pinning force (Fig. 1). Under the action of this
non--compensated component, proton fluxoids can be either
expelled from the core (line $1$ in Fig. 1), or aligned with the
neutron vortices (line $2$ and $2'$ in Fig. 1).

\begin{figure}
\begin{center}
\centering\includegraphics[height=9cm]{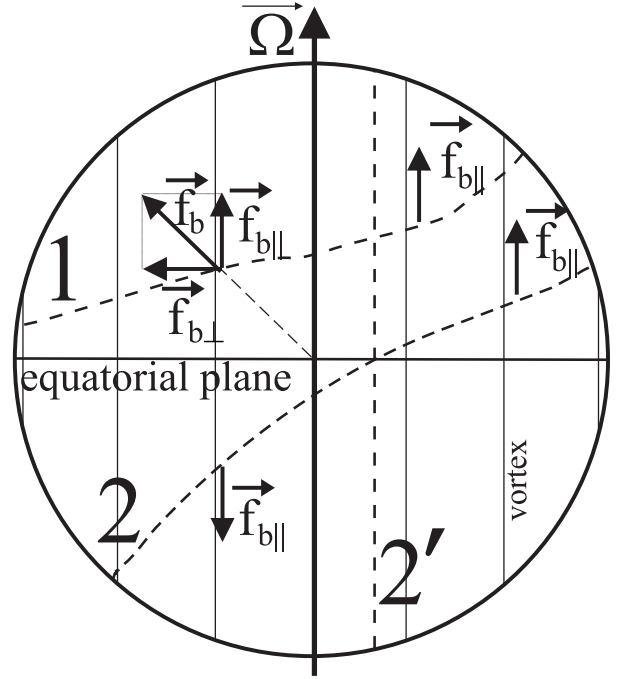}

\caption{The superfluid core of a NS rotates with the angular
velocity $\Omega$ by forming array of neutron vortices (shown in
the figure by thin vertical lines). Fluxoids are shown by dashed
lines.  The fluxoid 1, which does not intersect the equatorial
plane of the NS, will be expelled from the core along the
vortices under the action of noncompensated component of the
buoyancy force $\vec{f}_{\rm b ||}$, while its component
$\vec{f}_{\rm b \perp}$ will be compensated by the vortex acting
force. The fluxoid 2, which intersects the equatorial plane, will
become aligned with the vortices into position $2'$. The
timescales of both processes are the same.} \label{msp_fig1}
\end{center}
\end{figure}

\begin{figure*}
\centering\includegraphics[height=7.5cm]{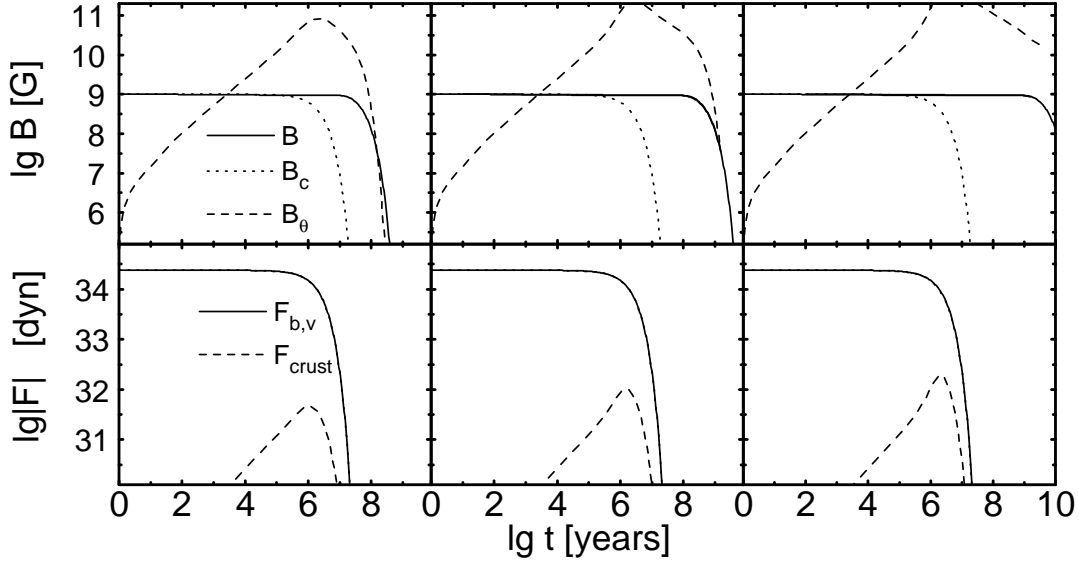}
\caption{The top panels show the evolution of MF strengths in the
core ($B_{\rm c}$), at the surface ($B$), and at magnetic equator
at the crust-core interface ($B_\theta$). The bottom panels show
forces, acting on the fluxoids in the core of MSP. Left, middle
and right columns correspond to $Q=0.1, 0.01$ and $0.001$. }
\label{fig:msp_fig2}
\end{figure*}

\noindent  The total flux of the Poynting vector through the
crust--core interface is equal to the power of forces,  acting
upon the fluxoids in the core (for details see KG00):
\begin{equation}
\sum_{\rm fluxoids}\int (\vec{f}_{\rm b ||}+\vec{f}_{\rm v})\cdot
\vec{v}_{\rm p}{{\rm d}l}= -\frac{c}{4\pi}\int\limits_{S_{\rm
core}}\left[\vec{E} \times \vec{B} \right] \cdot d\vec{S}_{\rm
core} , \label{main_eq}
\end{equation}
where the integral on the l.h.s. is taken  over the length of a
fluxoid, the summation runs over all fluxoids, and the
integration on the r.h.s. is performed over the crust--core
interface. For simplicity we assume that $f_{\rm b ||}=f_{\rm b}$
(this simplification leads to an underestimation of the time of
expulsion in the order of unity), that the number density of
fluxoids is uniform, and calculate all quantities at the density
of the crust--core interface. Thus we estimated the integral on
the l.h.s. of Eq.\ (\ref{main_eq}) as $4/3\cdot(f_{\rm b}+f_{\rm
v}) v_{\rm p} N_{\rm p} R_{\rm c}$ (see KG00), where $N_{\rm
p}=\pi B_{\rm c} R_{\rm c}^2/\Phi_0$ is the total number of
fluxoids and $B_{\rm c}$ the strength of the mean core MF. Note,
that Eq.\ (\ref{main_eq}) contains no force exerted by neutron
vortices, and describes the motion of fluxoids along vortices.

\noindent We assume the MF to be axisymmetric, poloidal, and
dipolar outside the NS. Thus, one can introduce the
vector--potential $A=(0,0,A_\phi)$, where $A_\phi=S(r,t) \sin
(\theta) /r=B_0 R^2 s(r,t) \sin (\theta) /r$, $s$ is  Stokes'
stream function, normalized to $B_0 R^2$, $B_0$ and $R$ are the
initial surface MF strength and radius of the NS, respectively.
The integral on the r.h.s. of Eq.\ (\ref{main_eq}) is given by
\begin{equation}
\frac{c}{4\pi}\int\limits_{S_{\rm core}}\left[\vec{E} \times
\vec{B} \right]
                           \cdot         d\vec{S}_{\rm core} =
\frac{ B_{0}^2 R^4}{6} \frac{\partial s(R_{\rm c},t)} {\partial t}
\frac{\partial s(R_{\rm c},t)} {\partial r}. \label{Pf_eq}
\end{equation}
We introduce the total core forces $F_{\rm b,v}=f_{\rm b,v} \cdot
4R_{\rm c}/3 \cdot N_p$, and the ``crustal'' force, which arises
when the roots of the fluxoids are moved along the crust--core
interface:

\begin{equation}
F_{\rm crust}=
       \frac{B_{0}^2 R^4}{6} \frac{1}{v_{\rm p}}
       \frac{\partial s(R_{\rm c},t)} {\partial t}
       \frac{\partial s(R_{\rm c},t)} {\partial r}.
\label{F_c_eq}
\end{equation}
Eq.\ (\ref{main_eq}) can be rewritten as
\begin{equation}
F_{\rm b}+F_{\rm v}(v_{\rm p})+F_{\rm crust}(v_{\rm p})=0.
\label{Forces_eq}
\end{equation}
The ``crustal'' force $F_{\rm crust}$ depends on $v_{\rm p}$ in a
rather complicated way, since both ${\partial s(R_{\rm c},t)} /
{\partial t}$ and ${\partial s(R_{\rm c},t)} / {\partial r}$
depend on $v_{\rm p}$. The evolution of the homogeneous MF within
the core is governed by $v_{\rm p}$ too (for details see KG00).
$F_{\rm crust}$ might be found by solving the induction equation
in the NS crust $(R_{\rm c}<r<R)$:
\begin{equation}
\frac{\partial s}{\partial t} = \frac{c^2}{4\pi\sigma}
\left(\frac{\partial^2 s} {\partial r^2} - \frac{2s}{r^2}\right),
\end{equation}
where $\sigma$ is the conductivity of NS crust. The crustal
conductivity is determined by collisions of the electrons on
impurities and phonons. For the phonon conductivity we use the
numerical data given by Itoh et al. (1993), for the impurity
conductivity we apply the analytical expression derived by
Yakovlev \& Urpin (1980). The magnitude of the impurity
conductivity  is determined by the impurity parameter $Q$, whose
value is highly uncertain and has to be varied. The impurity
conductivity dominates in the inner crust, where currents,
generated by the flux expulsion, are expected to be localized.

\noindent The boundary conditions are given by
$s(R,t)/R=-\partial s(R,t) / \partial r$ at the surface which
joins the MF inside the NS with the dipolar field in the vacuum
outside, and by Eq.\ (\ref{Forces_eq}) at the crust--core
interface. The initial condition is set by $B_0$ and the initial
$s$--profile in the NS.

\section{Results}

\noindent In Fig.2 we show the evolution of the MF strengths and
of the forces acting on the fluxoids for $B_0=B_{\rm c0}=10^9$ G
and an initial spin period $P_0=0.001$ s. The results in the
left, middle and right column are obtained for an impurity
parameter $Q=0.1, 0.01$ and $0.001$, respectively. The
computations were performed for a NS model, based on the
Friedman-Pandharipande equation of state (EOS), with NS mass
$M=1.4 M_{\odot}$, radius $R=10.4$ km, and thickness of the crust
$\Delta R=934$ m (Van Riper 1991).

\noindent The velocity of the fluxoids remains almost constant ($
\approx 7 \cdot 10^{-9}$ cm/s) during the MSP evolution for all
values of $Q$ considered here. This  is because in Eq.
(\ref{Forces_eq}) $F_{\rm crust}$ can be neglected in comparison
with $F_{\rm b}$ and $F_{\rm v}$ during the whole evolution. It is
seen from Fig. 2, that  $F_{\rm crust}\la 0.01(F_{\rm b},F_{\rm
v})$ and the fluxoid velocity is determined only by the balance of
buoyancy and drag forces. The expulsion timescale of the MF from
the core, $R_{\rm c}/v_{\rm p}$, is $\la 10^7$ years for all
values of $Q$ in the crust, whereas the decay timescale of the
surface MF is dependent on $Q$, and can be estimated by $10^7/Q$
years for the given EOS. Note, that the timescale of the
alignment must be in the same order of magnitude as the expulsion
timescale.

\noindent An interesting feature is the generation of a strong
$\theta$--component of the MF at the crust--core boundary.
Depending on $Q$ it may exceed the surface field strength by up
to three orders of magnitude when the flux expulsion proceeds.
This effect reflects the induction of currents close to the
crust--core interface, which are the stronger the longer the
characteristic decay time in the crust is.

\noindent We have performed similar computations for $B_0=10^8$
and $10^{10}$ G and values of the impurity parameter in the
interval $10^{-4}<Q<0.1$. We obtained the same results: the
expulsion timescale of the flux from the core is $\la 10^7$
years, the timescale of decay of the surface MF is  $10^7/Q$
years. That means that the expulsion timescale does not depend
on  the MF strength and $Q$ if $B_0 \le 10^{10}$ G. This is
because both $F_{\rm b}$ and  $F_{\rm v}$ are proportional to the
core MF strength, while $F_{\rm crust}$ is negligible. Our
estimate for the flux expulsion timescale in case of MF strengths,
characteristic for  MSPs, is consistent with that of Jones
(1987). Note, however, that for standard  pulsars  with $B \sim
10^{12}-10^{13}$ G, the expulsion timescale is dependent on the
MF strength. This is because $F_{\rm crust}$, which is
proportional to the $B_{\rm c}^2$, is then the dominating force,
which hampers the expulsion of the flux from the core of canonical
pulsars (KG00) and determines the expulsion timescale, balancing
$F_{\rm b}$. Though our model simplifies the real magnetic
configuration, we expect that it gives the correct estimate for
the expulsion timescale {\it along} the rotational axis ($10^7$
years). In contradiction, the expulsion {\it across} that axis
occurs on the spin--down timescale, which for MSPs exceeds $10^9$
years.

\section{Discussion}
If in MSPs the dissipation timescale of  crustal currents is less
than $10^9$ years, their surface MF would vanish. Alternatively,
the surface MF of MSPs can be aligned with their rotational axis
on timescales of  $< 10^9$ years. Both effects would make MSPs
unobservable as radiopulsars. The observed non--aligned
long--living surface MFs of MSPs can be sustained only if the
dissipation timescale of currents in the NS crust is $\ga 10^9$
years. Because of large differences in the thickness of the
crust, for a soft EOS the decay timescale of the crustal MF is
roughly given by $10^6/Q$ years, while for stiff EOS that
relation reads $10^8/Q$ years (Urpin \& Konenkov, 1997). Thus, for
a soft EOS and a thin crust the impurity parameter must be as low
as $Q \la 10^{-3}$, while for a stiff EOS with a thick crust $Q$
can be much higher ($Q \la 0.1$).

\noindent  The models of flux expulsion, developed by DCC, KG00,
KG01, and Jahan-Miri (2000) for standard pulsars, give an
inadequate low velocity of fluxoids at the late evolutionary
stages, resulting in the appearance of long--living residual MF
components. The main reason is that the equations of balance of
forces, which determine the velocity of fluxoids  (see Eq. (31) in
DCC, Eq. (11) in  Jahan-Miri 2000, Eq. (6) in  KG00), are written
in scalar form, whereas the forces are vectors. These equations
describe quantitatively only  the evolution of the component of
the fluxoid velocity directed perpendicular to the vortices. We
now rediscuss the papers KG00, KG01 and show, that this is a good
approximation for two evolutionary stages. I) In relatively young
($t<10^4$ years) standard radiopulsars  with $B\sim
10^{12}-10^{13}$ G this velocity component is determined by the
balance of vortex acting force $F_{\rm n}$ and the drag force
$F_{\rm v}$, and is about  $10^{-8}-10^{-7}$ cm/s (KG00). The
force exerted by fast outward moving neutron vortices exceeds the
buoyancy force. Thus we expect that the component of $v_{\rm p}$
perpendicular to the rotational axis would also exceed the
parallel component of $v_{\rm p}$.  II) Later, when  $F_{\rm n}
\ll (F_{\rm b}, F_{\rm crust})$, $F_{\rm b}$ is balanced by
$F_{\rm crust}$ thus determining the expulsion timescale. Then,
the fluxoids are almost not affected by the vortices when they are
moving towards the crust--core interface across them. At this
stage the bulk of the flux is expelled from the core, and we
conclude that the timescale of expulsion was estimated correctly.

\noindent  When almost all magnetic flux is expelled  from the
core, the vortex acting force is balanced by the perpendicular
component of buoyancy, thus hampering further flux expulsion {\it
across} the vortices. However, expulsion may proceed {\it along}
the vortices on a timescale of $\sim 10^7$ years. The appearance
of the residual MF in the above cited papers seems to be a
qualitatively wrong result. We do not expect any unaligned
magnetic flux to remain in the cores of old  ($10^9-10^{10}$
years) NSs with low ($B \le 10^{10}$ G) surface MFs.

\section{Conclusion}

The basic idea of this paper is that the motion of proton fluxoids
parallel to the neutron vortices is not restricted by the pinning
force, since the displacement of (even pinned) fluxoid parallel
to the vortex does not change the pinning energy. This approach
is basically different from those which consider the occurrence
of residual fields by the balance of buoyancy and vortex acting
forces at the late stage of NS evolution.\\

\noindent The velocity of fluxoids in the superfluid core of the
NS with low MF ($B \le 10^{10}$ G) is calculated. It has been
shown, that the fluxoids escape from the core or become aligned
with the neutron vortices on a timescale of $10^7$ years,
determined by the balance of the buoyancy and drag forces. This
timescale is independent of the MF strength and of the crustal
conductivity. The surface MF of the MSPs can be maintained on the
timescale of $\ga 10^9$ years only if the dissipation timescale
of the crustal currents is also $\ga 10^9$ years.  Therefore, an
estimate of the  impurity parameter $Q$ is possible. Once the NS
core matter behaves according a soft EOS this gives a limit of
$Q<0.001$, for a stiff EOS $Q<0.1$, counteracting the slower
field decay in a thicker NS crust. Because of the relatively weak
MF in MSPs neither the spin--down, nor the  ``crustal'' force
play an important role for the flux expulsion.\\

\noindent Many uncertainties affect the estimation of $Q$, both
for the crust of isolated and accreting neutron stars. However,
it is very unlikely that the impurity parameter is as small as
$10^{-3}$. De Blasio (1998, 2000) argues that $Q$ is relatively
large in that layers of the crust of isolated NSs, where the
chemical composition changes; in average the impurity parameter
should be larger than $10^{-3}$. The calculations of the chemical
composition and impurity content of the crusts of accreting NSs
show that $Q$ can be even as high as $\sim 100$ (Schatz et al.
1999). Since it is very likely that MSPs went through a phase of
intense accretion, this indicates -- under reserve that the
suppositions of our model are fulfilled -- that at least not too
soft EOS should describe the state of NS core matter. Such a
conclusion has been drawn also from the observation of kilohertz
quasi-periodic oscillations in several low mass X-ray binaries
(Klu\'zniak 1998) as well as from the consideration of the
magneto--rotational and thermal evolution of isolated NSs with
crustal MFs (Urpin \& Konenkov 1997; Page et al. 2000).

\noindent Clearly, the scenario will be qualitatively changed
when a portion of the core's volume remain in the normal state or
the forces acting on the fluxoid (Eq. (2)) have to be modificated
due to collective effects. \\

\begin{acknowledgements}

The authors are grateful to an anonymous referee for useful
comments. A discussion with Andreas Reisenegger is gratefully
acknowledged. The work of D.K. was supported by a scholarship of
the Alexander von Humbold--Stiftung, and by RFBR grant
00-02-04011.

\end{acknowledgements}

\end{document}